\documentclass[twocolumn,aps,prl]{revtex4-1}  
\usepackage[pdftex]{graphicx}
\usepackage{commath}
\usepackage{color}
\usepackage{lineno}
\usepackage{amsmath,amssymb}

\begin{document}
\title{Beam Energy Dependence of Jet-Quenching Effects in Au+Au Collisions at \\$\sqrt{s_{_{ \mathrm{NN}}}}$ = 7.7, 11.5, 14.5, 19.6, 27, 39, and 62.4\,GeV}


\author{
L.~Adamczyk$^{1}$,
J.~R.~Adams$^{29}$,
J.~K.~Adkins$^{19}$,
G.~Agakishiev$^{17}$,
M.~M.~Aggarwal$^{31}$,
Z.~Ahammed$^{51}$,
N.~N.~Ajitanand$^{40}$,
I.~Alekseev$^{15,26}$,
D.~M.~Anderson$^{42}$,
R.~Aoyama$^{46}$
% ...
}

\address{$^{1}$AGH University of Science and Technology, FPACS, Cracow 30-059, Poland}
% ...
\address{$^{15}$Alikhanov Institute for Theoretical and Experimental Physics, Moscow 117218, Russia}
% ...
\address{$^{17}$Joint Institute for Nuclear Research, Dubna, 141 980, Russia}
% ...
\address{$^{19}$University of Kentucky, Lexington, Kentucky 40506-0055}
% ...
\address{$^{26}$National Research Nuclear University MEPhI, Moscow 115409, Russia}
% ...
\address{$^{29}$Ohio State University, Columbus, Ohio 43210}
% ...
\address{$^{31}$Panjab University, Chandigarh 160014, India}
% ...
\address{$^{40}$State University of New York, Stony Brook, New York 11794}
% ...
\address{$^{42}$Texas A\&M University, College Station, Texas 77843}
% ...
\address{$^{46}$University of Tsukuba, Tsukuba, Ibaraki, Japan,305-8571}
% ...
\address{$^{51}$Variable Energy Cyclotron Centre, Kolkata 700064, India}
% ...

\begin{abstract}
We report measurements of the nuclear modification factor, $R_{ \mathrm{CP}}$, for charged hadrons as well as identified $\pi^{+(-)}$, $K^{+(-)}$, and $p(\overline{p})$ for Au+Au collision energies of $\sqrt{s_{_{ \mathrm{NN}}}}$ = 7.7, 11.5, 14.5, 19.6, 27, 39, and 62.4\,GeV.  We observe a clear high-$p_{\mathrm{T}}$ net suppression in central collisions at 62.4\,GeV for charged hadrons which evolves smoothly to a large net enhancement at lower energies.  This trend is driven by the evolution of the pion spectra, but is also very similar for the kaon spectra.   While the magnitude of the proton $R_{ \mathrm{CP}}$ at high $p_{\mathrm{T}}$ does depend on collision energy, neither the proton nor the anti-proton $R_{ \mathrm{CP}}$ at high $p_{\mathrm{T}}$ exhibit net suppression at any energy.  A study of how the binary collision scaled high-$p_{\mathrm{T}}$ yield evolves with centrality reveals a non-monotonic shape that is consistent with the idea that jet-quenching is increasing faster than the combined phenomena that lead to enhancement.\\

\end{abstract}

\maketitle

Evidence has been presented that high-energy heavy-ion collisions form a dense, nearly perfect, strongly-interacting, deconfined partonic liquid called quark-gluon plasma (QGP) \cite{Arsene:2004fa,Back:2004je,Adams:2005dq,Adcox:2004mh}.  This state of matter is thought to have dominated the universe prior to the hadron epoch \cite{Fromerth:2012fe}.  Quantifying the properties of the QGP is necessary for describing the QCD phase diagram \cite{Gyulassy:2004zy}, as well as constraining parameters in cosmological models that describe the evolution of the universe along a trajectory through the QCD phase diagram \cite{McInnes:2015hga}.  Just as the universe followed a particular trajectory through the QCD phase diagram, so do high-energy nuclear collisions.  The particular path for each collision depends on collision energy.  High-energy heavy-ion collisions form media with low initial baryon chemical potentials ($\mu_{\mathrm{B}}$) that are expected to remain low throughout their evolution.  This means that the trajectory passes through the region where a smooth crossover is predicted by theory \cite{Aoki:2006we,Borsanyi:2010bp}.  Lower collision energies have been shown to produce higher $\mu_{\mathrm{B}}$ \cite{Cleymans:2005xv,Andronic:2008gu}.  A first order phase transition is predicted at sufficiently high $\mu_{\mathrm{B}}$ \cite{Ejiri:2008xt, PhysRevC.79.015202} which would mean the existence of a critical end point \cite{Stephanov:2004wx}.  A beam energy scan (BES) program at the Relativistic Heavy-Ion Collider (RHIC) was proposed to further explore the QCD phase diagram, including a search for the critical point, and to demonstrate that signatures for QGP formation turn off at sufficiently low collision energies \cite{Aggarwal:2010cw}.  The STAR collaboration has recently published moments of net-proton and net-charge fluctuations in the BES as part of its critical point search \cite{Adamczyk:2013dal,Adamczyk:2014fia} with no evidence for the critical point within current uncertainties.  The future BES II program at RHIC will increase the acceptance and reduce the uncertainties for these measurements.  Implicit in the interpretation of these analyses was the requirement that a QGP be formed in the collisions at energies whose trajectories through the QCD phase diagram would pass near the critical point. Analyses are being carried out to determine at what collision energies signatures of QGP formation vanish.  Already published is the beam-energy dependence of charge separation along the magnetic field in Au+Au collisions with results consistent with a model featuring chiral symmetry restoration down to $\sqrt{s_{_{ \mathrm{NN}}}}$ = 11.5\,GeV \cite{Adamczyk:2014mzf}.  In another study, the third harmonic of azimuthal correlations was measured as a function of collision energy and the number of participating nucleons ($\langle N_{\mathrm{part}} \rangle$) \cite{Adamczyk:2016exq}.  Models have shown that the third harmonic is sensitive to the low viscosity of the QGP phase \cite{Muller:2007rs,Zajc:2007ey,Gale:2012rq}, and this measurement showed that the third harmonic persisted down to $\sqrt{s_{_{ \mathrm{NN}}}}$ = 7.7\,GeV for high-$\langle N_{\mathrm{part}} \rangle$ collisions.  Both of these low-$p_{\mathrm{T}}$ results are consistent with QGP being formed for $\sqrt{s_{_{ \mathrm{NN}}}}$ $\geq$ 11.5\,GeV so that the critical point would be directly accessible down to this collision energy.  While each of these measurements is compelling on their own, it is by constructing a body of independent measurements that we will gain confidence that the QGP is being formed at these low collision energies.  The measurements presented here focus on high-$p_{\mathrm{T}}$ probes of QGP formation: in particular, partonic energy loss, or jet-quenching.

High-$p_{\mathrm{T}}$ partons, the forebears of jets, are produced early in the collision, and while moving through QGP volume they lose energy via strong interactions.  This process is called jet-quenching \cite{Bjorken:1982tu,Wang:1991xy}.  Jet-quenching has contributions from collisional and radiative energy loss with strong force analogs to the processes described in chapters 13 and 14 respectively of Jackson's iconic text \cite{jackson1975classical}.  This would be expected to lead to a depletion, or suppression, of high-$p_{\mathrm{T}}$ hadrons in the final state.  

One method of observing this suppression is with the nuclear modification factor, $R_{ \mathrm{CP}}$, which is defined by Eq. (\ref{eq:Rcp}).  \begin{equation} \label{eq:Rcp}
 R_{ \mathrm{CP}} =  \frac{\langle N_{\mathrm{coll}}\rangle _{\mathrm{Peripheral}}}{\langle N_{\mathrm{coll}}\rangle _{\mathrm{Central}}}  \frac{(\frac{d^2N}{dp_{\mathrm{T}} d\eta})_{\mathrm{Central}}}{(\frac{d^2 N}{dp_{\mathrm{T}}d\eta})_{\mathrm{Peripheral}}} 
\end{equation}
 
Here, $N_{\mathrm{coll}}$ is the average number of binary collisions within a centrality bin and can be estimated using a Glauber Monte Carlo \cite{Miller:2007ri}.  If heavy-ion collisions were just a collection of $N_{\mathrm{coll}}$ independent $p$+$p$-like collisions, then $R_{ \mathrm{CP}}$ would be unity for the entire $p_{\mathrm{T}}$ range.  Effects that increase the number of particles per binary collision in central heavy-ion collisions relative to $p$+$p$ or peripheral collisions are collectively called enhancement effects and lead to $R_{ \mathrm{CP}}$ $\textgreater$ 1, while those that decrease the number of particles are collectively called suppression effects and lead to $R_{ \mathrm{CP}}$ $\textless$ 1.   This means that $R_{ \mathrm{CP}}$ can tell us whether enhancement or suppression effects are dominating, but not the magnitude of each separately.  Eq. (\ref{eq:Rcp}) compares the number of particles measured in small impact parameter (central) collisions where the mean pathlength through any produced medium would be longer, with large impact parameter (peripheral) collisions where the shorter in-medium pathlengths should result in less energy loss.  High-$p_{\mathrm{T}}$ suppression was observed at top Relativistic Heavy Ion Collider (RHIC) energies, $\sqrt{s_{_{ \mathrm{NN}}}}$ = 130 and 200\,GeV, soon after RHIC began running \cite{Arsene:2004fa,Back:2004je,Adams:2005dq,Adcox:2004mh} and later, at even higher energies, by experiments at the Large Hadron Collider (LHC) \cite{CMS:2012aa,Abelev:2014ypa}. 

High-$p_{\mathrm{T}}$ suppression is expected to vanish at low collision energies, where the energy density becomes too low to produce a sufficiently large and long-lived QGP.  Another effect that may lead toward suppression at the lower collision energies is the EMC effect, a suppression of per nucleon cross sections in heavier nuclei relative to lighter nuclei for Bjorken $x > 0.3$, first measured with deep inelastic scattering by the European Muon Collaboration (EMC) \cite{Aubert:1983xm}.  While their measurement was for an impact parameter averaged nuclear modification of the parton distribution function (nPDF), what we are interested in here is the impact parameter dependence of this effect \cite{Helenius:2012ny}. Experimentally quantifying this and other possible cold nuclear matter (CNM) effects that affect these measurements would require reference data for the BES, $p$+$p$ and $ p$($d$)+Au.

Several physical effects could enhance hadron production in specific kinematic ranges, concealing the turn-off of the suppression due to jet-quenching.  One such effect is the Cronin effect; a CNM effect first observed in asymmetric collisions between heavy and light nuclei, where an enhancement of high-$p_{\mathrm{T}}$ particles was measured rather than suppression \cite{Cronin:1974zm,Antreasyan:1978cw,Straub:1992xd}.  It has been demonstrated that the enhancement from the Cronin effect grows larger as the impact parameter is reduced \cite{Vitev200336,Accardi2004244}.  Other processes in heavy-ion collisions such as radial flow and particle coalescence may also cause enhancement \cite{Greco:2003mm}.  This is due to the effect of increasing particle momenta in a steeply falling spectra.  A larger shift of more abundant low-$p_{\mathrm{T}}$ particles to higher momenta in more central events --- such as from radial flow, pt-broadening, or coalescence --- would lead to an enhancement of the $R_{ \mathrm{CP}}$.  These enhancement effects would be expected to compete with jet-quenching, which shifts high-$p_{\mathrm{T}}$ particles toward lower momenta.  This means that measuring a nuclear modification factor to be greater than unity does not automatically lead us to conclude that a QGP is not formed.  Disentangling these competing effects may be accomplished with complementary measurements, such as event plane dependent nuclear modification factors \cite{Adare:2012wg}, or through other methods like the one developed in this letter. 

\begin{figure}
\begin{center}
\includegraphics[width=20pc]{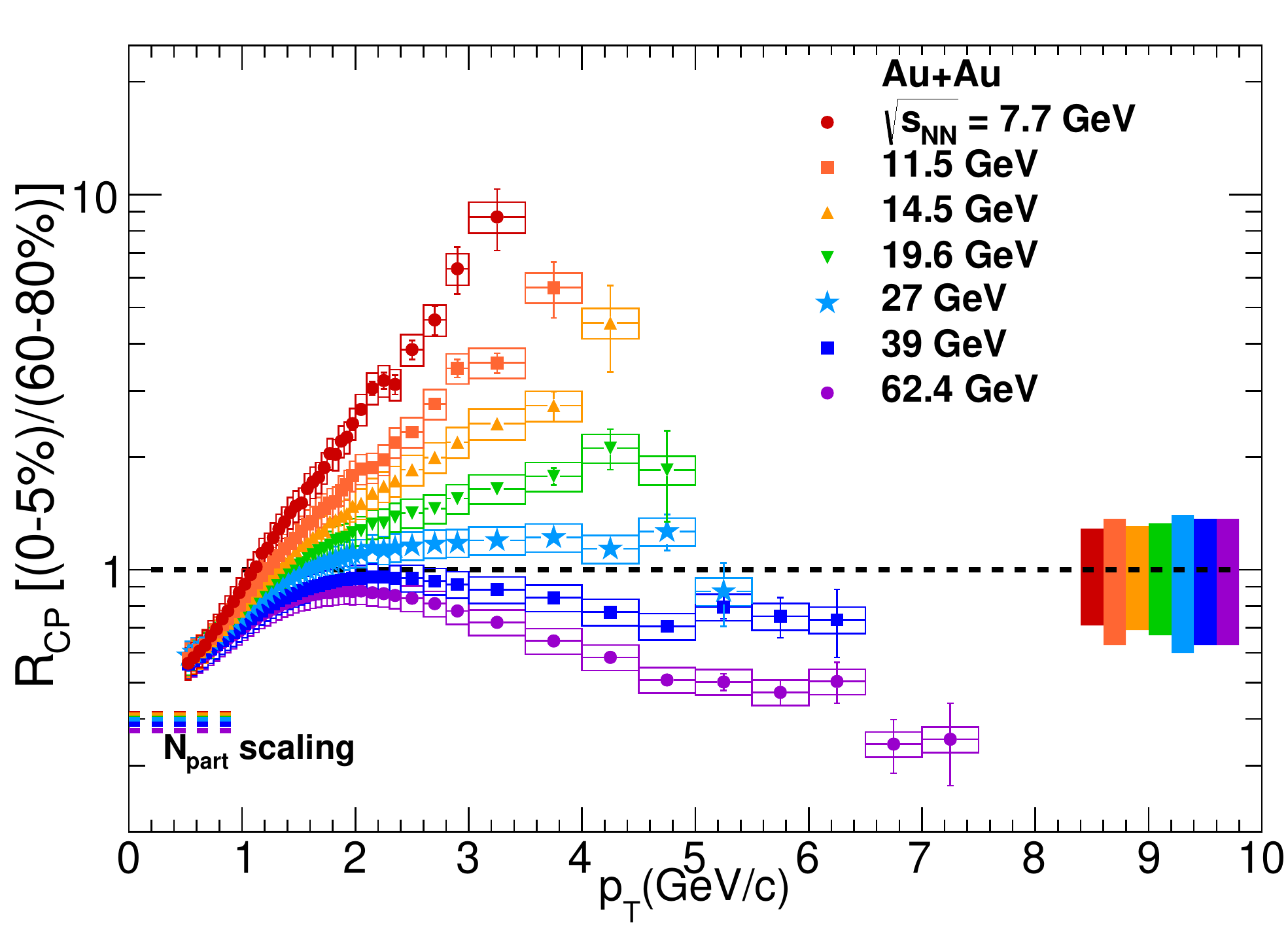}
\end{center}
\caption{\label{label}Charged hadron $R_{ \mathrm{CP}}$ for RHIC BES energies.  The uncertainty bands at unity on the right side of the plot correspond to the $p_{ \mathrm{T}}$ independent uncertainty in $N_{ \mathrm{coll}}$ scaling with the color in the band corresponding to the color of the data points for that energy.  The vertical uncertainty bars correspond to statistical uncertainties and the boxes to systematic uncertainties.}
\end{figure}

\begin{figure*}
\includegraphics[width=40pc]{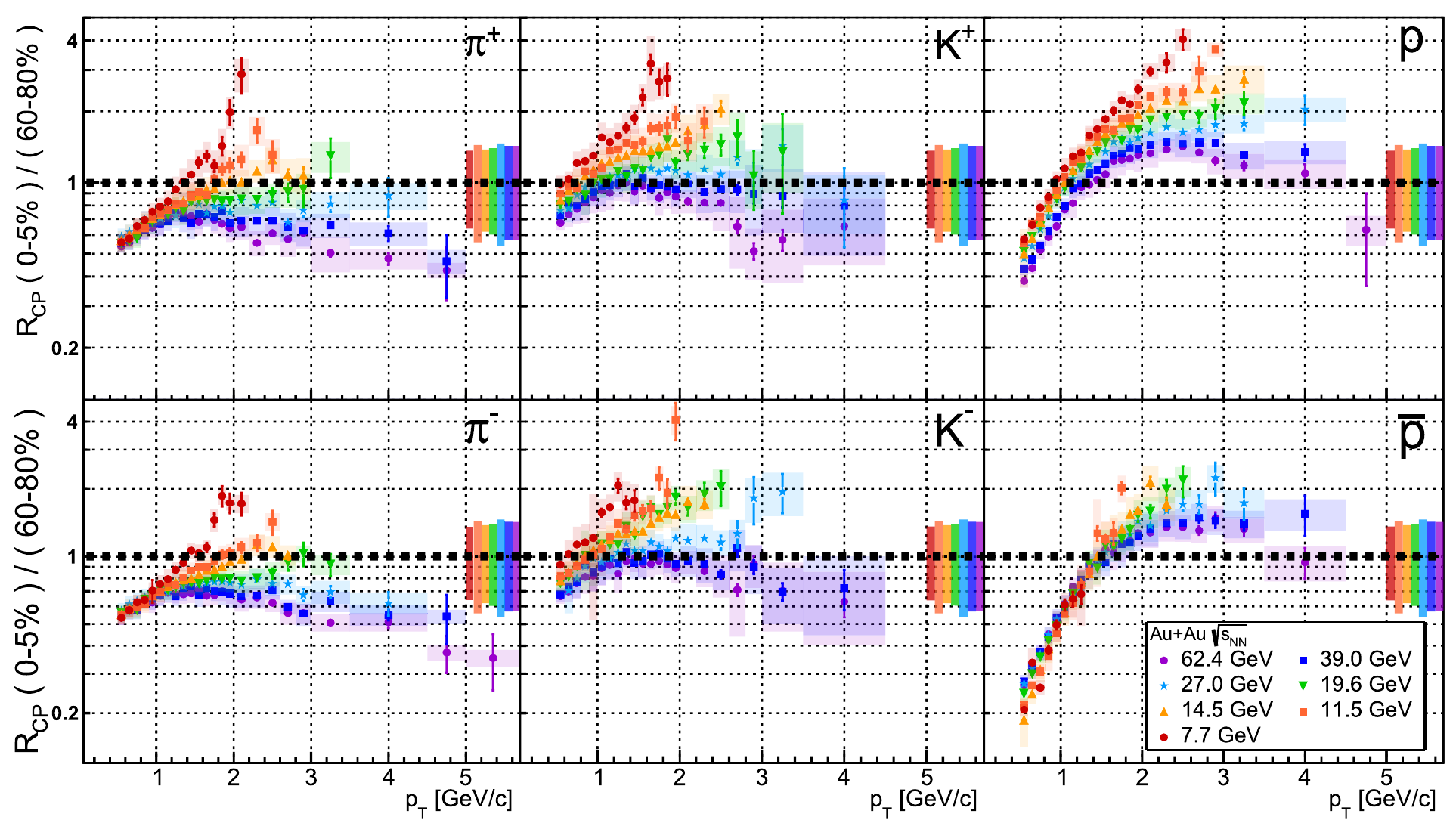}\hspace{2pc}%
\begin{minipage}[b]{34pc}\caption{\label{label3}Identified particle $R_{ \mathrm{CP}}$ for RHIC BES energies.  The colored shaded boxes describe the point-to-point systematic uncertainties. The uncertainty bands at unity on the right side of the plot correspond to the $p_{ \mathrm{T}}$ independent uncertainty in $N_{ \mathrm{coll}}$ scaling with the color in the band corresponding to the color of the data points for that energy.  }
\end{minipage}
\end{figure*}

\begin{figure}
\begin{center}
\includegraphics[width=20pc]{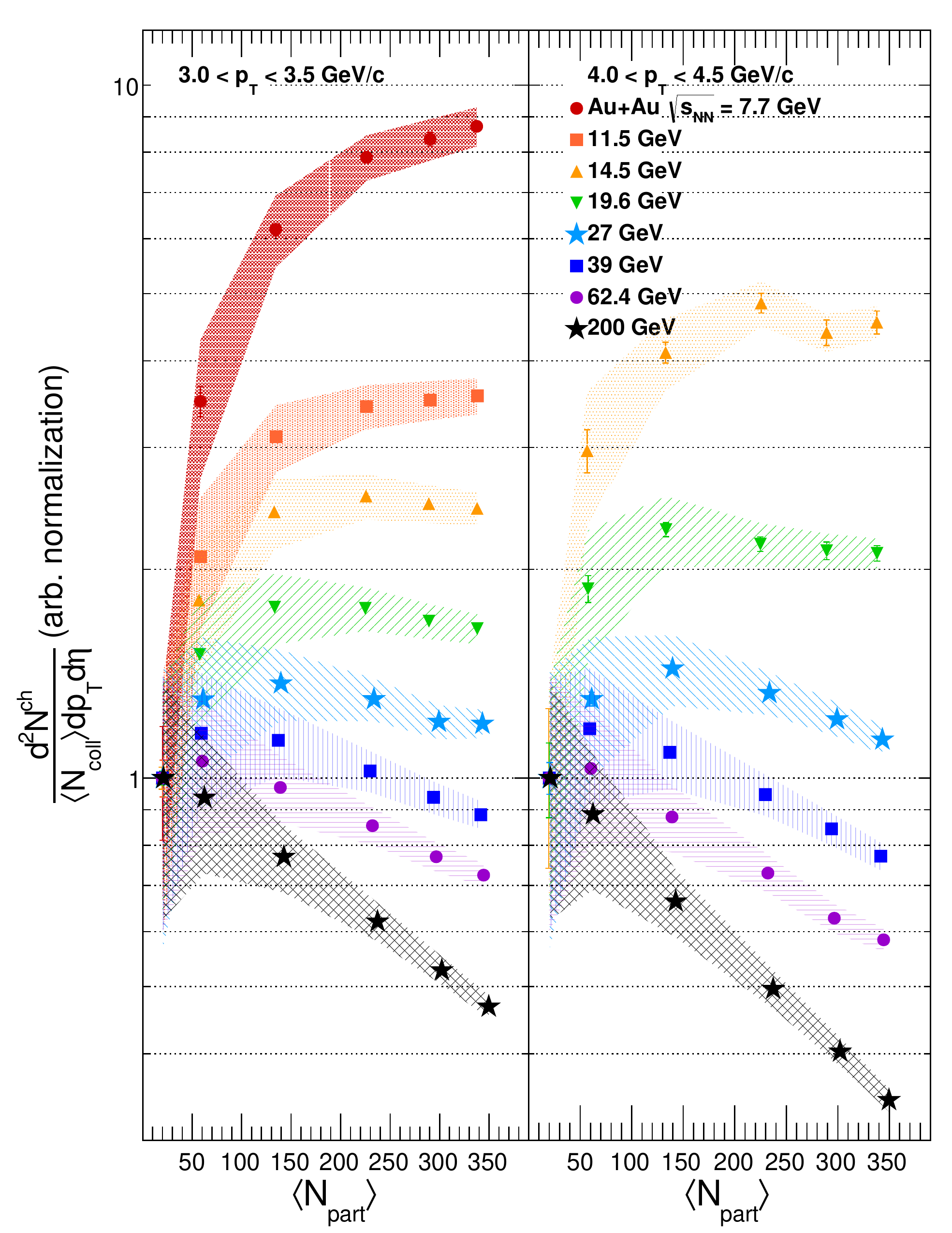}
\end{center}
\caption{\label{label2}Charged hadron $Y(\langle N_{\mathrm{part}}\rangle )$ for two ranges of $p_{\mathrm{T}}$.  Statistical uncertainty bars are included, mostly smaller than point size, as well as shaded bands to indicate systematic uncertainties.  The centrality bins from left to right correspond to 60-80\%, 40-60\%, 20-40\%, 10-20\%, 5-10\%, and 0-5\%.}
\end{figure}

In this letter we report measurements sensitive to partonic energy-loss, performed by the STAR experiment at several energies below $\sqrt{s_{_{ \mathrm{NN}}}}$ = 200\,GeV.  The data for this analysis were collected in the 2010, 2011, and 2014 RHIC runs by the STAR detector \cite{Ackermann:2002ad}.  STAR is a large acceptance detector whose tracking and particle identification for this analysis were provided by its Time Projection Chamber (TPC) \cite{Anderson:2003ur} and Time-of-Flight (TOF) \cite{Llope:2012zz} detectors.  These detectors lie within a 0.5\,T magnetic field that is used to bend the paths of the charged particles traversing it for momentum determination.  Minimum bias triggered events were selected by requiring coincident signals at forward and backward rapidities in the Vertex Position Detectors (VPD) \cite{Llope:2014nva} with a signal at mid-rapidity in the TOF.  The VPDs also provide the start time for the TOF system, with the TOF's total timing resolution below 100\,ps \cite{Llope:2012zz}.  Centrality was determined by the charged multiplicity at mid-rapidity in the TPC.  The only correction to the charged multiplicity comes from the dependence of the tracking efficiency on the collision's vertex position in the TPC.  Events were selected if their position in the beam direction was within $30$\,cm of the TPC's center and if their transverse vertex position was within 1\,cm of the mean transverse position for all events.  Tracks were accepted if their distance of closest approach to the reconstructed vertex position was less than 1\,cm, they had greater than 15 points measured in the TPC out of a maximum of 45, and the number of points used in track reconstruction divided by the number of possible points was greater than 0.52 in order to prevent split tracks.  The $p_{\mathrm{T}}$ and species dependent tracking efficiencies in the TPC were determined by propagating Monte Carlo tracks through a simulation of STAR and embedding them into real events for each energy and centrality \cite{Anderson:2003ur}.  The charged hadron tracking efficiency was then taken as the weighted average of the fits to the single species efficiencies with the weights provided by fits to the corrected spectra of each species.  This method allowed for extrapolation of charged hadron efficiencies to higher $p_{\mathrm{T}}$ than the single species spectra could be identified.  The efficiencies were constant as a function of $p_{\mathrm{T}}$ in the extrapolated region, which limited the impact from the extrapolation on the systematic uncertainties.  Daughters from weak decay feed-down were removed from all spectra.  The corrections for absorption and feed-down were determined by passing events generated in UrQMD \cite{Bass:1998ca} through a STAR detector simulation.  Charged tracks in $\abs{\eta}$ $<$ 0.5 and identified particles with  $\abs{y}$ $<$ 0.25 were accepted for this analysis.  Particle identification was performed using both energy loss in the TPC ($dE/dx$) and time-of-flight information ($1/\beta$).  

The overall scaling systematic uncertainty for the $R_{\mathrm{CP}}$ measurements is dominated by the determination of $N_{\mathrm{coll}}$ and the total cross section, which is driven by trigger inefficiency and vertex reconstruction efficiency in peripheral events.  Point-to-point systematic uncertainties arise from the determination of the single particle efficiency (5\% for the $p_{\mathrm{T}}$ range studied here), momentum resolution (2\%), and feed-down ($p_{\mathrm{T}}$ and centrality dependent with a range of 4-7\%). These systematic uncertainties are highly correlated as a function of centrality and $p_{\mathrm{T}}$ with the different sources of uncertainty added in quadrature.   Point-to-point systematic uncertainties for identified species have an additional contribution from uncertainties in particle identification that grow larger as the $dE/dx$ and $1/\beta$ bands for the different species merge at higher momenta.  The contribution from particle identification to the systematic uncertainties is small (1-3\%) at low $p_{\mathrm{T}}$ and large (up to 9\%) at high $p_{\mathrm{T}}$.

Figure 1 shows the $\sqrt{s_{_{ \mathrm{NN}}}}$ and $p_{\mathrm{T}}$ dependence of charged hadron $R_{\mathrm{CP}}$ constructed with data from (0-5)\% and (60-80)\% event centralities.  The $R_{\mathrm{CP}}$ is found to be lowest at the highest beam energy studied, and increases progressively from a suppression regime at 62.4\,GeV to a pronounced enhancement at the lowest beam energies.  This enhancement may have contributions from Cronin type interactions \cite{Cronin:1974zm,Antreasyan:1978cw,Straub:1992xd}, radial flow \cite{Greco:2003mm}, and the relative dominance of coalescence versus fragmentation for hadronization \cite{Greco:2003mm}.  Number of participant nucleons ($\langle N_{\mathrm{part}} \rangle$) scaling, which is expected to be more appropriate for bulk particle production at lower $p_{\mathrm{T}}$, is shown on the y-axis. This plot demonstrates the turn-off of net suppression for high-$p_{\mathrm{T}}$\ hadrons produced in central collisions relative to those produced in peripheral collisions.  This meets, for this signature of QGP formation, one of the goals of the BES \cite{Aggarwal:2010cw}.  Figure 1 clearly demonstrates that enhancement effects become very large at lower energies. Therefore in order to identify at what collision energy QGP is formed, more sensitive observables are required.  The next step is to look for more sensitive probes that could reveal potential evidences of jet-quenching at lower collision energies.

In order to extract $R_{ \mathrm{CP}}$ for identified hadrons, the particles rapidity density ($dN/dy$) is used in Eq. (\ref{eq:Rcp}).  Figure 2 shows $R_{ \mathrm{CP}}$ as a function of $p_{\mathrm{T}}$ for feed-down subtracted identified particles at different collision energies.  While net enhancement of high-$p_{\mathrm{T}}$ particles is observed at all energies for $p$ and $\overline{p}$, high-$p_{ \mathrm{T}}$ $\pi^{+(-)}$ are suppressed for both 39 and 62.4\,GeV, which drives the trends observed in charged hadrons.  $K^{+(-)}$ have similar energy dependence to $\pi^{+(-)}$, but show less net suppression.  The $R_{ \mathrm{CP}}$ of protons seems to turn over for the highest two energies.  The large suppression of low-$p_{ \mathrm{T}}$ $\overline{p}$ $R_{ \mathrm{CP}}$ is consistent with a picture of annihilation prior to kinetic freeze-out \cite{ShivaKumar:1994nk}.  Suppression in $R_{ \mathrm{CP}}$ of pions persists to lower collision energies than that of charged hadrons; this is likely due to smaller enhancement from the Cronin effect, radial flow, and coalescence for pions than protons.  These measurement of $\pi^{+(-)}$ $R_{ \mathrm{CP}}$ are consistent with measurements of $\pi^{0}$ $R_{\mathrm{AA}}$ in Au+Au collisions at $\sqrt{s_{_{ \mathrm{NN}}}}$ $\geq$ 39\,GeV \cite{Adare:2012uk}, and with  $\pi^{0}$ $R_{\mathrm{CP}}$ in Pb+Pb collisions at $\sqrt{s_{_{ \mathrm{NN}}}}$ = 17.3\,GeV \cite{Aggarwal:2001gn}.  However, while earlier measurements demonstrated the disappearance of net suppression, the results presented here extend to lower collision energies where a strong net enhancement is observed.

A measurement of $R_{\mathrm{CP}}$ takes the ratio of $N_{\mathrm{coll}}$-scaled spectra from two different centralities \cite{sup}.  A new and more differential method to study jet-quenching is to look at how the $N_{\mathrm{coll}}$-scaled spectra trend with centrality for a high-$p_{\mathrm{T}}$ bin.  
\begin{equation} \label{eq:Y}
 Y(\langle N_{\mathrm{part}}\rangle ) =  \frac{\mathrm{1}}{\langle N_{\mathrm{coll}}\rangle }\frac{d^2N}{dp_{\mathrm{T}} d\eta}(\langle N_{\mathrm{part}}\rangle ) 
\end{equation}
This is equivalent to taking the numerator from $R_{\mathrm{CP}}$ and plotting it versus centrality so that the peripheral bin contents are in the first bin at low $\langle N_{\mathrm{part}} \rangle$ and the central bin's contents are in the last point at high $\langle N_{\mathrm{part}} \rangle$.  Examining the full centrality evolution allows for the disentanglement of whether the processes leading to enhancement increase faster or slower than the processes leading toward suppression as a function of centrality.  While both jet-quenching and enhancement effects increase in strength with increasing $\langle N_{\mathrm{part}} \rangle$, if there is a faster growth of quenching, it would manifest itself in decreasing $Y({\langle N_{\mathrm{part}}} \rangle)$ trends.  To simplify comparison of these centrality trends across all energies each distribution is normalized by the contents of its most peripheral bin.

Figure 3 shows the charged hadron yield per binary collision in two ranges of $p_{\mathrm{T}}$ as a function of $\langle N_{\mathrm{part}} \rangle$.  These results are shown for 3 $\textless$ $p_{ \mathrm{T}}$ $\textless$ 3.5\,GeV/c in the left panel of Fig. 3 and for 4 $\textless$ $p_{ \mathrm{T}}$ $\textless$ 4.5\,GeV/c in the right panel.  The left panel corresponds to the highest $p_{ \mathrm{T}}$ bin of the $\sqrt{s_{_{ \mathrm{NN}}}}$ = 7.7\,GeV data and the right panel is for the highest $p_{ \mathrm{T}}$ bin of the $\sqrt{s_{_{ \mathrm{NN}}}}$ = 14.5\,GeV data.  Similar results are obtained for all $p_{ \mathrm{T}}$ $\textgreater$ 2\,GeV/c but the kinematic reach is smaller for low $\sqrt{s_{_{ \mathrm{NN}}}}$.  The 200\,GeV points are from STAR data taken in 2010 and analyzed with the same procedure as the BES points.  The measurement of $Y(\langle N_{\mathrm{part}} \rangle)$ decreases monotonically for $\sqrt{s_{_{ \mathrm{NN}}}}$ = 200\,GeV with increasing $Y(\langle N_{\mathrm{part}} \rangle)$, as expected for stronger an increase of quenching effects with increasing collision centrality compared to the effects leading to enhancement.  The measurement of $Y(\langle N_{\mathrm{part}} \rangle)$  increases monotonically for 7.7 and 11.5\,GeV data meaning that enhancement effects increase faster than suppression effects as you go more central for these collision energies.  For the other collision energies enhancement effects increase faster than suppression effects at first, but as you go more central suppression effects begin to increase faster than enhancement effects and the (0-5)\% central scaled yields are suppressed relative to less central scaled yields.  For example, at 14.5 GeV it can be seen that enhancement increases faster than suppression for all centrality bins from 60-80\% down to 10-20\%. However in the two most central bins, 5-10\% and 0-5\%, suppression effects increase at a similar rate to enhancement effects.  In fact, if the systematic errors are taken to be 100\% correlated, which is reasonable over this range of centralities, then the (0-5)\% yields are significantly suppressed relative to less central yields.  This may be interpreted as medium-induced jet-quenching decreasing high-$p_{ \mathrm{T}}$ yields in central collisions at $\sqrt{s_{_{ \mathrm{NN}}}}$ $\gtrsim$ 14.5\,GeV.  As we move to higher energies we can see evidence for jet-quenching in less central collisions.  This does not exclude the possibility of QGP formation in the 7.7 and 11.5\,GeV datasets, but simply that enhancement effects increase faster than quenching effects for all centralities at these energies.  This hadronic dominance at lower energies is consistent with what was measured for other QGP signatures in the BES \cite{Adamczyk:2014mzf,Adamczyk:2013gv,Adamczyk:2016exq}.  

In summary, net high-$p_{ \mathrm{T}}$ suppression persists for charged hadron $R_{\mathrm{CP}}$ for $\sqrt{s_{_{ \mathrm{NN}}}}$ \textgreater 39\,GeV.  Partonic energy loss may still occur at lower $\sqrt{s_{_{ \mathrm{NN}}}}$ with Cronin-like enhancement competing with this suppression effect and so observables that may be less sensitive to enhancement effects are considered as well.  Mesons and baryons are observed to have different trends with the $R_{ \mathrm{CP}}$ of high-$p_{\mathrm{T}}$ baryons being enhanced at every energy in the RHIC BES.  This points toward pion $R_{ \mathrm{CP}}$ as a cleaner observable for medium induced jet-quenching with pion $R_{ \mathrm{CP}}$ suppressed for $\sqrt{s_{_{ \mathrm{NN}}}}$ \textgreater 27\,GeV.   Finally, using a new method developed in this paper where we study the scaled yield as a function of centrality, we have measured suppression of charged hadrons in 0-5\% centrality events relative to a centrality bin where enhancement effects have already begun to dominate (10-20\%) for $\sqrt{s_{_{ \mathrm{NN}}}}$ $\gtrsim$ 14.5\,GeV.   This high-$p_{ \mathrm{T}}$ result does not rule out the possibility that QGP is also formed in $\sqrt{s_{_{ \mathrm{NN}}}}$ $\textless$ 14.5\,GeV since it is only sensitive to whether suppression effects increase faster than enhancement effects with increasing $\langle N_{\mathrm{part}} \rangle$.  Instead it frames a consistent picture with previous measurements to support a model where QGP is formed in central collisions at $\sqrt{s_{_{ \mathrm{NN}}}}$ \textgreater\ 14.5\,GeV.  

We thank the RHIC Operations Group and RCF at BNL, the NERSC Center at LBNL, and the Open Science Grid consortium for providing resources and support. This work was supported in part by the Office of Nuclear Physics within the U.S. DOE Office of Science, the U.S. National Science Foundation, the Ministry of Education and Science of the Russian Federation, National Natural Science Foundation of China, Chinese Academy of Science, the Ministry of Science and Technology of China and the Chinese Ministry of Education, the National Research Foundation of Korea, GA and MSMT of the Czech Republic, Department of Atomic Energy and Department of Science and Technology of the Government of India; the National Science Centre of Poland, National Research Foundation, the Ministry of Science, Education and Sports of the Republic of Croatia, RosAtom of Russia and German Bundesministerium fur Bildung, Wissenschaft, Forschung and Technologie (BMBF) and the Helmholtz Association.

\bibliographystyle{apsrev4-1}
\bibliography{rcp_1_1}

\end{document}